\documentclass[final]{aipproc}

\layoutstyle{8x11double}

\usepackage{graphicx}
\usepackage{framed}
\usepackage{indentfirst}
\usepackage{color}
\usepackage{natbib}
\usepackage{amsmath,amssymb}
\usepackage{subfigure}
\usepackage[margin=1in]{geometry}

\begin{document}

\title{\fontsize{18}{16.8} \selectfont Multiple-choice Assessment for Upper-division Electricity and Magnetism}


\author{\fontsize{14}{15.6} \selectfont Bethany R. Wilcox}{
  address={Department of Physics, University of Colorado, 390 UCB, Boulder, CO 80309}
  }

\author{\fontsize{14}{15.6} \selectfont Steven J. Pollock}{
  address={Department of Physics, University of Colorado, 390 UCB, Boulder, CO 80309}
  }

\begin{abstract}
The Colorado Upper-division Electrostatics (CUE) diagnostic was designed as an open-ended assessment in order to capture elements of student reasoning in upper-division electrostatics.  The diagnostic has been given for many semesters at several universities resulting in an extensive database of CUE responses.  To increase the utility and scalability of the assessment, we used this database along with research on students' difficulties to create a multiple-choice version.  The new version explores the viability of a novel test format where students select multiple responses and can receive partial credit based on the accuracy and consistency of their selections.  This format was selected with the goal of preserving insights afforded by the open-ended format while exploiting the logistical advantages of a multiple-choice assessment.  Here, we present examples of the questions and scoring of the multiple-choice CUE as well as initial analysis of the test's validity, item difficulty, discrimination, and overall consistency with the open-ended version.
\end{abstract}

\maketitle

\section{Introduction \& Motivation}

Historically, data collected from large-scale conceptual assessments, like the BEMA (Brief Electricity and Magnetism Assessment) \cite{Ding2006}, has help to drive course transformations and investigations of student learning in introductory physics \cite{Meltzer2012}.  Data from upper-division conceptual assessments, like the Colorado Upper-division Electrostatics (CUE) diagnostic, have similar potential.  The CUE is a free-response (FR) instrument that explicitly asks students for their reasoning on problems related to upper-division electrostatics.  Scores on the CUE correlate strongly with other measures of student learning such as overall course and BEMA score, and are sensitive to several different types of instruction (e.g., interactive engagement vs. traditional lecture) \cite{Chasteen2012}.

The CUE has been given for multiple semesters at several institutions; however, it requires a complex rubric and significant training to grade consistently.  This severely limits its potential as a large-scale assessment tool like the multiple-choice (MC) instruments used at the introductory level.  If the CUE is to be used as a tool by a wide range of faculty, it must be adapted to a more easily graded format without sacrificing its ability to provide a meaningful measure of students' conceptual learning in upper-division electrostatics.  

Previous work by Lin and Singh \cite{Lin2011} has looked at the difference between physics questions in MC and FR formats.  They crafted several MC questions where the distractors were based on common student difficulties and scores were weighted to reflect different levels of understanding. Comparing scores on MC and FR versions of these questions, they found that average scores on the two formats did not differ significantly, and that both formats had similar discrimination.  

Using student solutions from previous semesters to help construct distractors, we have developed a multiple-choice version of the CUE that allows students to receive partial credit depending on the accuracy and consistency of their solutions.  This paper describes the development and scoring of the MC CUE as well as a preliminary comparison of scores on the MC and FR versions.

\vspace*{-14pt}
\section{The Multiple-choice CUE}
\vspace*{-4pt}

\begin{figure}
  \begin{minipage}{1\linewidth}
  \begin{framed}
    \vspace{-1mm}\flushleft \fontsize{9}{10.8}\selectfont {\bf Q3} - A neutral non-conducting cube as below, with $\rho(z)=kz$.  Find $\vec{E}$ or V at point P, where P is off-axis, at a distance {\bf 50a} from the cube.\\
    \vspace{1mm} Select only one: {\bf The easiest method would be ...}

   \begin{minipage}{0.52\linewidth}
      \vspace{1mm}\flushleft
      A. Direct Integration\\
      B. Gauss's Law\\
      C. Separation of Variables\\
      D. Multipole Expansion\\
      E. Ampere's Law\\
      F. Method of Images\\
      G. Superposition\\
      H. None of these
   \end{minipage}
   \begin{minipage}{0.45\linewidth}    
      \begin{center}
        \includegraphics[width=22mm, height=30mm]{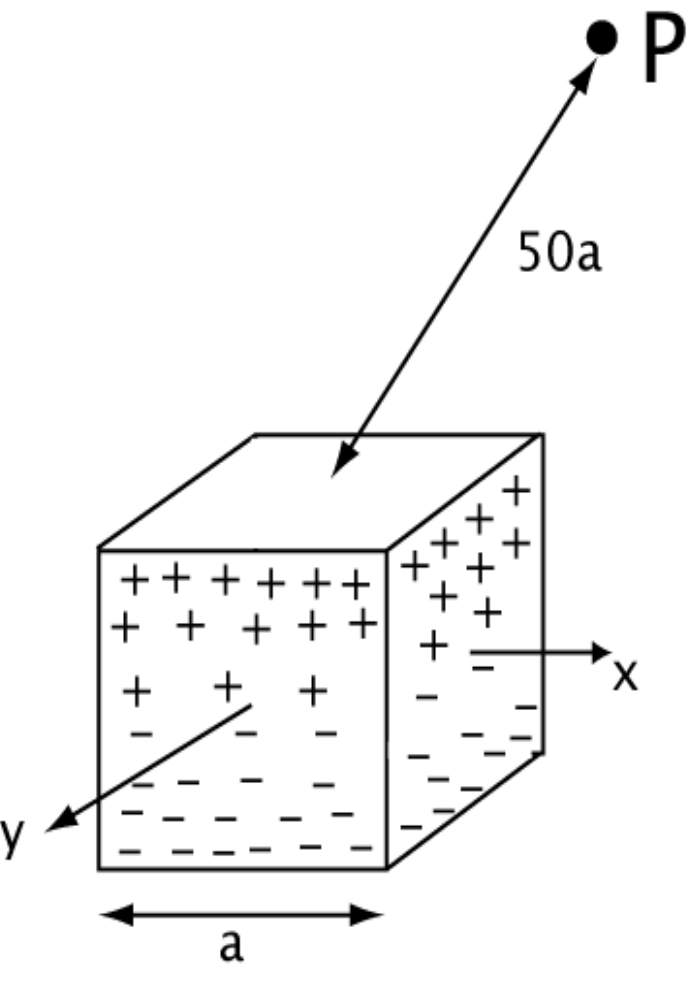}
      \end{center}
   \end{minipage}
     \flushleft{{\bf because ...} (select {\bf ALL} that support your method choice)\\
      a. $\Box$ you can calculate $\vec{E}$ or V using the integral form of Coulomb's Law\\
      b. $\Box$ the cube will look like a dipole; approximate with $\vec{E}$ or V for an ideal dipole\\
      c. $\Box$ symmetry allows you to calculate $\vec{E}$ using a cubical Gaussian surface\\
      d. $\Box$ symmetry allows you to calculate $\vec{E}$ using a spherical Gaussian surface\\
      e. $\Box$ the observation point is far from the cube\\
      f. $\Box$ there is not appropriate symmetry to use other methods\\
      g. $\Box$ $\nabla^2 V = 0$ outside the cube and you can solve for V using Fourier Series} \vspace{-4pt}
  \end{framed}
  \end{minipage}
\caption{A sample item from the MC CUE.} \label{fig:MCexample}
\end{figure}

{\bf Adapting the Questions:} We developed the MC CUE by closely examining student responses to the FR version.  It quickly became clear that the standard MC format of a single, unambiguously correct answer with four to six tempting distractors would be insufficient to capture the range of student reasoning.  For many of the CUE questions, a completely correct justification requires that the student connect several distinct ideas.  Many students give partially correct justifications that are missing one or more key elements.  To accommodate this trend, we provided a set of different reasoning elements of which students can select all that support their answer (see Fig.\ \ref{fig:MCexample}).  Reasoning elements were taken from student responses to the FR version and may be correct, incorrect, or irrelevant in the context of a specific question.  Full credit requires that a student select all (and only) the reasoning elements that together form a complete justification; however, they can receive partial credit for selecting some, but not all, of the necessary elements.   The wording of the question prompts was adjusted only when necessary to accommodate the new format.  A sample item from the MC CUE is given in Fig.\ \ref{fig:MCexample}.  The boxes next to each reasoning element are intended to resemble `check all' boxes that the students are accustomed to seeing online.  

{\bf Scoring:} The MC format also allows for considerable flexibility in terms of scoring.  A FR test requires significant time to regrade with a different rubric, but it is relatively simple to produce scores on the MC CUE using different grading schemes.  Here, we discuss two potential scoring rubrics.  The first rubric (R1) attempts to preserve the straightforward scoring of a standard multiple choice test.  The grading scheme in R1 awards correct responses full points and incorrect responses no points.  On the reasoning portion of the questions (see Fig.\ref{fig:MCexample}), points are distributed between all the appropriate reasoning elements weighted by the relative importance of each element to justifying the correct response.  

The second rubric (R2) attempts to replicate the nuanced grading used to score the FR CUE \cite{Chasteen2012}.  The grading scheme for R2 awards full points for correct responses, but also awards partial credit for selecting methods that are possible, even if they are not easy.  R2 also awards points for reasoning elements that are consistent with the choice of method.   For example, for the item shown in Fig.\ \ref{fig:MCexample}, the Multipole Expansion is the easiest method; however, it is also possible to use Direct Integration.  R2 awards students who select method `A' some partial credit and additional partial credit for selecting the consistent reasoning element, `a'.  R2 also subtracts points from students with reasoning elements that are inconsistent with their choice of method.  Differences between scores obtained using R1 and R2 are discussed below.


{\bf Expert Validation:} The FR CUE was designed to align with explicit learning goals developed by collaborative faculty working groups.  The instrument was also reviewed by physics experts to establish that the physics content was accurate, clear, and valued \cite{Chasteen2012}.  Because the MC CUE has the same questions, the validity of the physics content is, to a large extent, already established.  However, the operationalization of this content has changed significantly in the new format.  We solicited and received feedback from eight content experts at multiple institutions with experience teaching upper-division physics.  Small modifications were made to the phrasing of several items as a result of this feedback.  Overall, the expert reviewers expressed enthusiasm for the MC CUE and offered no critiques that questioned the overall validity of the new format.  

{\bf Student Validation:} Think-aloud validation interviews are crucial to ensure that students interpret the questions, instructions, and distractors on the MC CUE as intended, particularly because the `select ALL that apply' format may not be familiar.  To date, we have performed three interviews with a 7 question subset of the full 16 question MC CUE and six interviews with the full instrument.  All interview participants had completed an upper-division electrostatics course one to four weeks prior to the interview with final course scores ranging from A to C.  During these interviews, students were asked to externalize their reasoning.  At the end, the interviewer probed the students in more detail where it was unclear why they selected or rejected certain distractors.  Students were also asked to articulate what procedure they associated with each of the `Method' options.  Minor wording changes were made to the reasoning elements on the MC CUE as a result of these interviews.  

A concern raised by one faculty reviewer was that students who did not know how to start a problem might figure out the correct approach by examining the given reasoning elements.  We did observe instances in the interviews where students would explicitly refer to the reasoning elements in order to inform their choice of method.  However, this technique seemed most useful to students with higher course scores, and, in all such cases, the student provided additional reasoning that clearly demonstrated their understanding of the correct method.  Alternatively, some students in the interviews were led down the wrong path by focusing on an inappropriate reasoning element.  This suggests that students using the reasoning elements to figure out the answer does not result in a significant inflation of scores.

\vspace*{-14pt}
\section{Preliminary Results}
\vspace*{-4pt}

\begin{figure}
\includegraphics{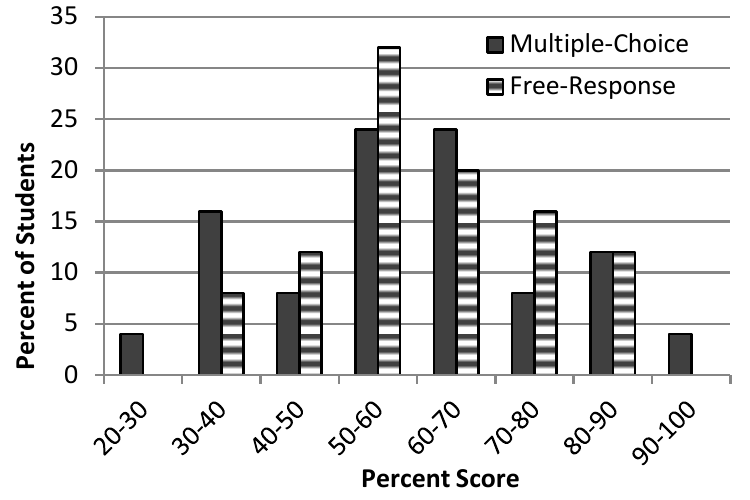}
\caption{Distributions of scores on the MC and FR CUE (N=25 in each) for the E\&M 1 course at CU.  There is no statistically significant difference between the distributions (Student's t-test, p = 0.7). }\label{fig:comparison}
\end{figure}

The MC CUE was administered for the first time in an upper-division electrostatics course (N=62) at the University of Colorado Boulder (CU).  This course, Electricity and Magnetism 1 (E\&M 1), is the first of a two semester sequence that typically covers the first six chapters of Griffith's text \cite{Griffiths1999}.  The course was taught by a PER faculty member (SJP) and incorporated a number of materials designed to promote interactive engagement, such as in-class tutorials and clicker questions \cite{Chasteen2012trans}.  

A first-order goal with the development of the MC CUE was to achieve a meaningful level of agreement between the scores on the new MC and well-established FR versions \cite{Chasteen2012}.  To make this comparison, we gave half of the E\&M 1 students the MC version and half the FR version.  The two groups were randomly selected but matched based on average midterm exam score.  Attendance on the day of the diagnostic was typical and 25 students took each version of the CUE.    While the size of this class was unusually large (even for CU), the conclusions that can be drawn from the analysis presented in the remainder of this section are still limited by low-N.  Additional testing in E\&M courses with different instructors and at additional institutions is needed to determine the robustness of these findings.  

Using the more nuanced grading rubric (R2), the average score on the MC version, 58.9 $\pm$ 3.6 \%, was not significantly different from the average on the FR version,  60.5 $\pm$ 2.8 \% (Student's t-test, p = 0.7).  Score distributions for both versions  (Fig.\ \ref{fig:comparison}) were nearly normal (Anderson-Darling test, p > 0.85) and had similar variances (Brown-Forsythe test, p > 0.9).  Additionally, the stop time of each student was recorded and, on average, students spent a comparable amount of time on the MC version, 35 $\pm$ 2 min, as on the FR version, 37 $\pm$ 2 min.  The difference in time was not significant (p = 0.4).  

The average score on the MC version from the simpler grading scheme (R1) was 56.1 $\pm$ 3.7 \%.  This score does not differ statistically from the average scores on either the FR version (p = 0.4) or the R2 grading scheme (p = 0.6).  However, the more nuanced grading of R2 was designed to better reflect subtle differences in levels of understanding between students, and it may be that the size of this initial sample was not large enough to highlight these differences.  Because it more closely matches the FR rubric, the remainder of the analysis exclusively utilizes scores from the R2 grading scheme.  

{\bf Criterion Validity:} Another important property of the MC CUE overall is how well its scores correlate with other, related measures of student understanding.  The most straightforward comparison is with course exam scores.  Students in E\&M 1 took two midterm exams and one final exam.  The MC CUE scores correlate strongly with aggregate exam scores (r=0.78, p < 0.05).  For comparison, the correlation for the 25 students who took the FR version was also high (r = 0.75, p < 0.05).  Similarly the scores for both versions are strongly correlated with total course score (MC: r = 0.78, FR: r = 0.68).  These correlations are somewhat higher than the r $\approx$ 0.5 reported previously for the FR CUE \cite{Chasteen2012}.  This may be because the instructor (SJP) was involved in the test's development and its questions align well with his philosophy on teaching and learning goals.  Stronger claims will be possible as we continue to collect MC CUE data.  

\begin{figure}
\includegraphics{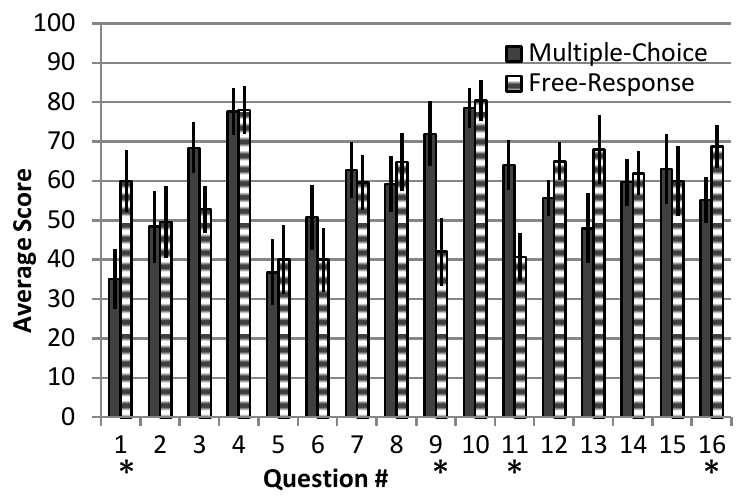}
\caption{Average scores on each item on the CUE.  Statistically significant differences between the MC and FR versions are indicated by an asterisk.  Error bars represent 1$\sigma$ error. All questions are available from Ref.\ \cite{contact}}\label{fig:difficulty}
\end{figure}

{\bf Item Difficulty:} In addition to looking at the overall performance of students on the MC and FR versions of the CUE, it is important to examine their performance on individual items.  Fig.\ \ref{fig:difficulty} shows the average scores on both versions for each question.  As shown in the figure, differences between the scores are significant for 4 of 16 items (Mann-Whitney U-test, p < 0.05).  It is tempting to assume that the MC version would be easier; however, this is only the case for 2 of the 4 (Q9 \& 11), where interviews suggest that  students performed better on the MC because the answers to these items are more easily recognized than generated (e.g., boundary conditions on E and V).  For one of the other two (Q1), students were more likely to select an inappropriate solution method on the MC version.  The difference in scores on remaining question (Q16) may have been due to the constraints of the MC version.  Students no longer receive credit for working through the math to show their answer and instead must justify it conceptually.  

{\bf Item Discrimination:} It is also valuable to determine how well performance on each item compares to performance on the rest of the test (i.e., how well each item discriminates between high and low performing students).  Item-test correlations were between 0.34 and 0.63 for all items on the MC CUE with the exception of Q8 (r = 0.29), Q9 (r = 0.19), and Q13 (r = 0.12).  For this population there were no statistically significant differences between the item-test correlations for the MC and FR versions; however, previous publications on the CUE report item-test correlations of 0.5 or greater for all items \cite{Chasteen2012}.  This suggests that the lower correlations on some items observed here is a feature more of this specific course than of the instrument in general.  A common criteria for acceptable item-test correlations is r $\geq$ 0.2 \cite{Ding2006}; however, for N = 25 correlation coefficients less than 0.31 are not statistically significant.  Additional data will be necessary to make robust statements about item discrimination.

As an additional measure of the discriminatory power of the MC CUE, we calculate Ferguson's Delta \cite{Ding2006}.  Ferguson's Delta is a measure of how well distributed scores are over the full range of possible scores.  It can take on values between [0,1] and any value greater than 0.9 indicates good discriminatory power \cite{Ding2006}. For this student population, Ferguson's Delta for both the MC and FR versions of the CUE is 0.95.  This is similar to the previously reported FR value of 0.99 \cite{Chasteen2012}.

{\bf Internal Consistency:} The consistency of scores on individual items is also important.  To examine this, we calculate Cronbach's Alpha for the test as a whole.  While the CUE violates the underlying assumption that the test measures a single construct \cite{Chasteen2012}, Cronbach's Alpha will still provide a conservative measure of the internal consistency of the instrument.  Using the point value of each item to calculate alpha, we find $\alpha$ = 0.80, which is also the commonly accepted cutoff for a good value \cite{Graham2006}; thus the MC CUE has an acceptable level of internal consistency. Again, this is consistent with the value of 0.82 reported for the FR CUE \cite{Chasteen2012}.

\vspace*{-14pt}
\section{Concluding Remarks}
\vspace*{-4pt}

We have created a multiple-choice version of an existing upper-division conceptual assessment, the CUE.  Using student responses to the original free-response version of the instrument, we crafted multiple-choice distractors which reflected common student ideas.  This new version utilizes a novel approach to multiple-choice questions that allows students to select multiple reasoning elements in order to construct a complete justification for their answers.  By awarding points based on the accuracy and consistency of students' selections, this assessment has the potential to produce scores that reflect subtle differences between students' understanding.  

Student interviews, expert feedback, and preliminary analysis of scores on the MC CUE from one upper-division electrostatics course have all yielded promising measures of the validity and reliability of the instrument as a whole.  Scores on the new multiple-choice version of the CUE also show a high degree of agreement with scores on the original free-response version.  Given this, the logistical advantages of a multiple-choice instrument make the new format considerably more viable as a tool for large-scale implementation.  

One of the primary goals of the CUE's original creators was to gain insight into the details of common student difficulties \cite{Chasteen2012}.  We argue that the `select ALL that apply' MC format still allows us to gain useful insight into student thinking.  By identifying patterns in students' method and reasoning selections, it is possible to examine the nature and consistency of student ideas.  For example, the majority of our students tend to select reasoning elements that are consistent (or at least, not inconsistent) with their choice of method; however, students' answers are less consistent between subparts of the same question.  Students' method selection helps us to pinpoint topics where the students struggle, while their reasoning selections give us insight in to the nature of the difficulty.   We are continuing to explore how much insight into student thinking can be extracted from the MC CUE.  

Further testing of the MC CUE with additional instructors and at other institutions will be necessary to establish the robustness of the findings presented here.  This will also increase the statistical power of these analysis to pinpoint differences between the MC and FR versions of the CUE and help to establish the validity and reliability of the new format for different student populations.  Additional data collection will also be needed to determine if the MC CUE retains the FR version's sensitivity to differences in pedagogy.  See Ref.\ \cite{contact} for more information on reviewing or administering the MC CUE.

\vspace*{-14pt}
\begin{theacknowledgments}
\vspace*{-4pt}
Particular thanks to the PER@C group, the faculty who reviewed the CUE, and M.D. Caballero for his statistical expertise.
This work was funded by NSF-CCLI Grant DUE-1023028 and a National Science Foundation Graduate Research Fellowship under Grant No.1548433.
\end{theacknowledgments}

\vspace*{-10pt}
\bibliographystyle{aipproc}   
\bibliography{MC-CUE-PERC-2013-refs}

\end{document}